\documentclass[aps,prl,reprint,showpacs,groupedaddress]{revtex4-1}
\usepackage{amssymb}
\usepackage{graphicx}
\usepackage{amsmath}

\usepackage{color}

\begin{document}

\setcounter{page}{1}

\title{Stabilization and self-passivation of symmetrical grain boundaries by mirror symmetry breaking}

\author{Ji-Sang Park}
\affiliation{Thomas Young Centre and Department of Materials, Imperial College London, Exhibition Road, London SW7 2AZ, UK}
\email[]{ji-sang.park@imperial.ac.uk}

\bibliographystyle{apsrev4-1}

\date{\today}

\begin{abstract}

While computational methods for the study of point defects in materials have received significant attention, methods for the investigation of grain boundaries require further development.
In this study, we applied a genetic algorithm to find the atomic structure of grain boundaries in semiconductor materials with given Miller indices and investigated their electronic structure.
Our study of the $\Sigma$3 (112) grain boundary in CdTe shows that the stability of grain boundaries can be greatly enhanced by breaking mirror symmetry, locally or globally.
The resulting grain boundary can be electrically less harmful because the origin of the defect levels are removed from the middle of the band gap and the grain boundaries can serve as a channel for electron extraction.

\end{abstract}

\maketitle

The change in physical properties resulting from a broken symmetry is certainly one of the most important subjects of condensed matter and material physics \cite{pantelides1978electronic,hirth1982theory,gleiter1982structure,smith1990theory,rurali2010colloquium,kang2013origin,major2016grain,hellman2017interface,park2018point}.
Translational symmetries can be broken in various ways by the presence of defects. 
Considerably more efforts have been devoted to understanding point defects than extended defects such as dislocations and grain boundaries. 
This is partly because the control of point defects is vital for achieving efficient semiconductor devices (e.g. transistors, light emitting diodes, and solar cells) by tuning the Fermi level and suppressing trap-assisted recombinations \cite{chadi1988theory,fahey1989point,freysoldt2014first,yang2016review,park2018point}.
Efforts to investigate the role of point defects have now become mature enough to allow for the automation of the calculation of point defects \cite{goyal2017computational,broberg2018pycdt}.

Extended defects play a critical role in materials as exemplified in studies of a wide range of materials such as III-nitrides \cite{nakamura1998roles}, graphene \cite{grantab2010anomalous}, metals \cite{lu2016stabilizing}, and thin-film solar cells \cite{li2014grain}, which have been investigated by both experimental and computational studies.
Electron backscattering diffraction (EBSD) can be used to identify microstructural properties of grain boundaries, providing useful information to select grain boundaries of interest \cite{humphreys2001review,moseley2015recombination}.
Further atomistic details of grain boundaries can be investigated by Transmission Electron Microscopy (TEM)  \cite{kim2001nonstoichiometry,yan2003structure,klie2005enhanced,sun2013creating,li2014grain}. 
Atomistic calculations have been successfully used to construct a three-dimensional atomic structure of grain boundaries from two-dimensional images obtained from TEM experiments  \cite{cheng2017anomalous,taha2017grain,cash2018local,zhu2018role,chavez2018molecular,liebscher2018strain,kim2018first}.
Learning from experimental characterizations of grain boundaries can provide computational models with physically meaningful constraints. This is especially important when modelling extended defects because the many degrees of freedom inherent in grain boundaries make them difficult to study, and these physical constraints reduce the configurational space.

\begin{figure}
\includegraphics[width=\columnwidth]{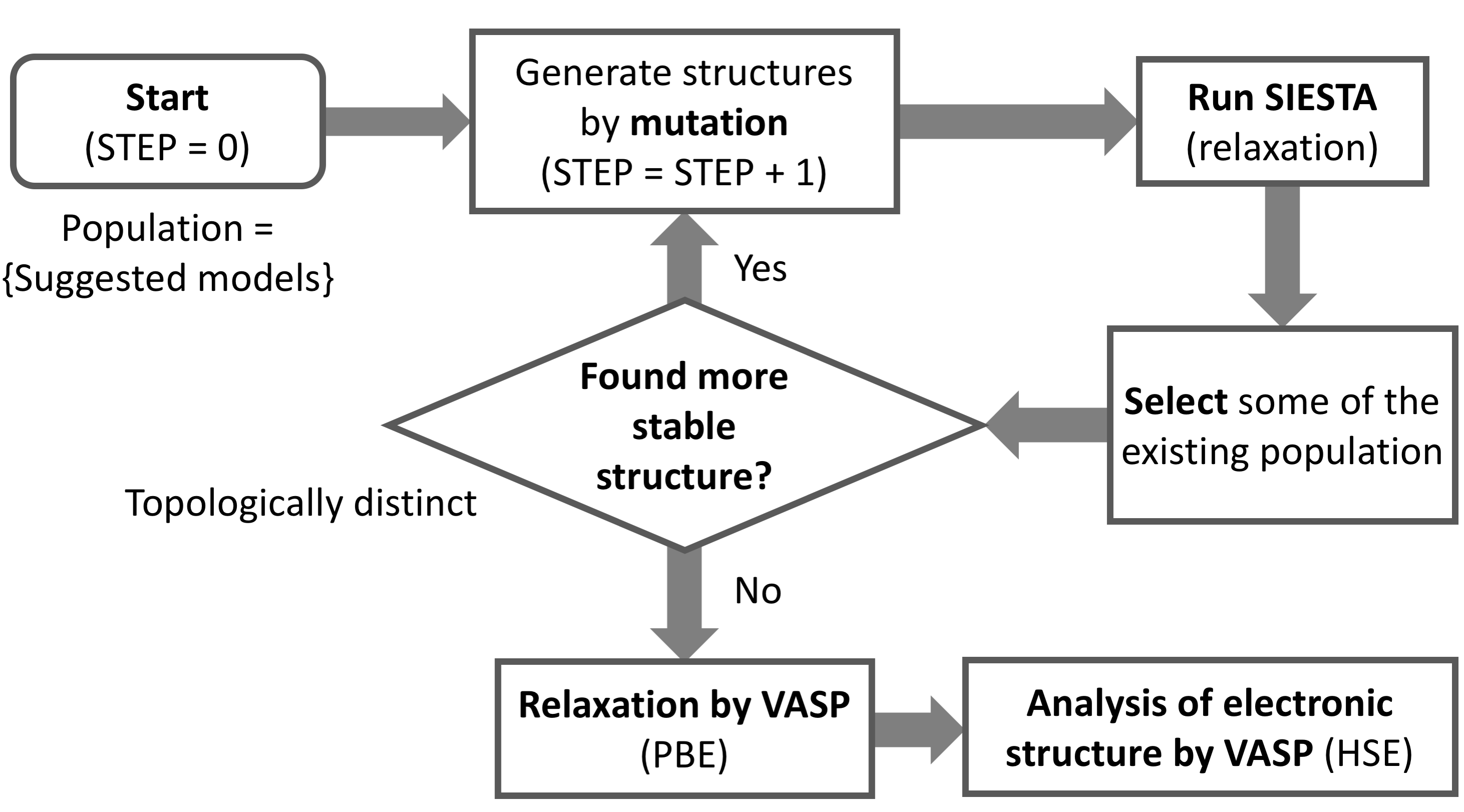}
\caption{\label{fig:1} Flowchart of the calculation. Mutation operators include the formation of Frenkel pairs and the relative shift of a grain with respect to another (rigid body translation). Initially, the previously suggested models were included in the population set.}
\end{figure}

In this study, we employed a computational strategy to obtain grain boundary atomic structures in semiconductors when minimal information is given. Here, we suggest an efficient strategy to stabilize and self-passivate grain boundaries by breaking a mirror symmetry at the boundary, as illustrated in Figure 1.
A genetic algorithm in conjunction with an inexpensive quantum mechanical calculation method was employed to find a stable grain boundary structure.
Some selected atomistic models were further relaxed and analyzed using a plane wave basis method.
Our calculations of grain boundaries in CdTe show that the extended defects can be self-passivated by the symmetry breaking, which causes the reconstructions and removes the defect levels from the middle of the gap.
Stabilization of the grain boundary by mirror symmetry breaking can be also examined in other systems.

Genetic algorithms require several operations to generate new children from parents. 
The rule of thumb we use is that an over-coordinated atom should be placed in a void region, which is less populated by atoms, and surrounded by atoms with opposite charges.
Therefore, when we generated new structures from the parent structures, we chose an over-coordinated atom and placed it at a new position while the distance from neighboring atoms is maximized (Frenkel-type defects, A$_{\mathrm{A}}$$\rightarrow$A$_i$+V$_{\mathrm{A}}$).
We also considered the parallel shifting of one of the grains, which breaks the mirror symmetry.
This makes the grain boundary a glide plane by definition because a glide plane consists of a reflection followed by a translation.
This operation is termed as rigid body translation (RBT) in the study of metal grain boundaries \cite{gleiter1982structure}.
When one of the grains is shifted in a parallel direction to the grain boundary plane, we attempted to find a structure maximizing the number of bonds.

Computational studies of grain boundaries in semiconductors are often performed using the Generalized Gradient Approximations (GGA) as the cell size along the direction normal to a boundary should be thick enough to reproduce bulk-like regions \cite{yan2003structure,park2015stability,lazebnykh2015ab,liu2016self}.
A slab geometry with a grain boundary is usually used to prevent potential charge transfer between the two boundaries in the supercell calculations, and at least $\sim$10 {\AA} thickness of the vacuum region should be considered.
The GGA calculations are computationally less heavy than the hybrid density functional theory (DFT) calculations, but the usage of the plane wave basis can be computationally heavy if our goal is to investigate several hundreds of grain boundary models.

\begin{table}
\begin{center}
\caption{Formation energy ($E_f$) of grain boundaries calculated by the PBE and the HSE06 exchange-correlation functional. $E_f$ of model A is set to 0 eV/nm$^2$. Atomic structure of grain boundaries is shown in Figure 2. NKRED was set to 2 in the HSE06 calculations.}
\begin{tabular}{ c c c c }
\hline\hline 
Model & $E_f$(PBE) (eV/nm$^2$) & $E_f$(HSE06) (eV/nm$^2$) & Ref. \\ \hline
A &  0 & 0 & \cite{yan2003structure} \\ \hline
B & -0.40 & -0.21& \cite{liu2016self} \\ \hline
C & 0.00  & 0.19 & \\ \hline
D & -1.12  & -0.76 & \\ \hline \hline
\end{tabular}
\end{center}
\end{table}

To make the high-throughput calculation of the grain boundaries feasible, we used the Spanish Initiative for Electronic Simulations with Thousands of Atoms (SIESTA) code, which is another quantum chemistry code based on the molecular orbital basis \cite{soler2002siesta}.
The relative total energy between grain boundary models is calculated using single-$\zeta$ basis and/or double-$\zeta$ basis set. 
Keeping the supercell size, the atomic structures were relaxed using conjugate-gradient methods but atoms close to the surfaces are fixed in position to maintain bulk-like regions.
Since we modeled CdTe grain boundaries using slab geometries, there are two CdTe surfaces in a supercell. Cd and Te atoms have 2 and 6 valence electrons, and thus each Cd and Te dangling bond has 0.5 e and 1.5 e, respectively. The Cd and Te dangling bonds were passivated by pseudo-hydrogen atoms which have 1.5 e and 0.5 e, respectively.
We note that the lattice constant was set to the experimental value of 6.48 {\AA} \cite{panicker1978cathodic,lalitha2004characterization}.

While the total energy was used as a figure of merit (i.e. fitness function) to screen grain boundary models, structurally identical models were removed from the population set to remove redundancy. For this aim, we constructed a matrix $b$ having the information of the bonding network in the grain boundary model.
An element of a matrix, $b_{ij}$, was set to 1 if the distance between atom $i$ and atom $j$ are equal to or lower than the threshold value. 
Otherwise, if two atoms are more than the threshold distance apart $b_{ij}$ was set to 0.
The determinant of the matrix $b$ in conjunction with the enthalpy was used to distinguish grain boundary models.

After each step, ten models were selected among the population in the previous step and the newly examined structures.
If we found a topologically different stable structure than the existing structures, we repeated the cycle to search configuration spaces.
When we could not find new structures after a few more steps, we terminated the iterations and some chosen models were relaxed again by the Vienna Ab initio Simulation Package (VASP) code \cite{kresse1996efficient}.
For the VASP calculation, we used the exchange-correlation functional parametrized by Perdew, Burke, and Ernzerhof (PBE) \cite{perdew1996generalized} and Projector-Augmented Wave (PAW) pseudo-potentials \cite{blochl1994projector}.
An energy cutoff of 400 eV was used, and the smallest spacing between \textit{k}-points was set to $\simeq$0.03$/${\AA}.

Nowadays mixing of the GGA functional with the Hartree-Fock exact exchange, which is known as the hybrid DFT method, is widely used for analyzing the electronic structure.
Among various functionals, the hybrid functional suggested by Heyd, Scuseria, and Ernzerhof (HSE06) was used in this study \cite{HSE}.
Hybrid DFT calculation with a dense \textit{k}-point mesh is computationally heavy. In order to reduce the HSE06 calculations to a feasible level, we performed the self-consistent-field (SCF) calculations using a reduced \textit{k}-point set for the Fock exchange potential by setting NKRED to 2 in the VASP code.

Among many reported grain boundaries, we chose $\Sigma$3 (112) grain boundaries in CdTe to examine our strategy because their atomic structure has been studied using both TEM and density functional theory (DFT) calculations \cite{yan2003structure,liu2016self}.
The atomic structure of selected $\Sigma$3 (112) grain boundary models is shown in Figure 2. 
Since both grains have the Miller index of \{112\}, the grain boundaries are categorized as symmetric tilt grain boundaries, also known as twin boundaries. 
One thing should be pointed out is that the detailed atomic structure does not necessarily have a mirror symmetry. If two grains have the same Miller index and two grains are not rotated to each other (i.e. the twist angle of 0$^\circ$), then their boundary is a symmetric tilt grain boundary by definition.

\begin{figure}
\includegraphics[width=\columnwidth]{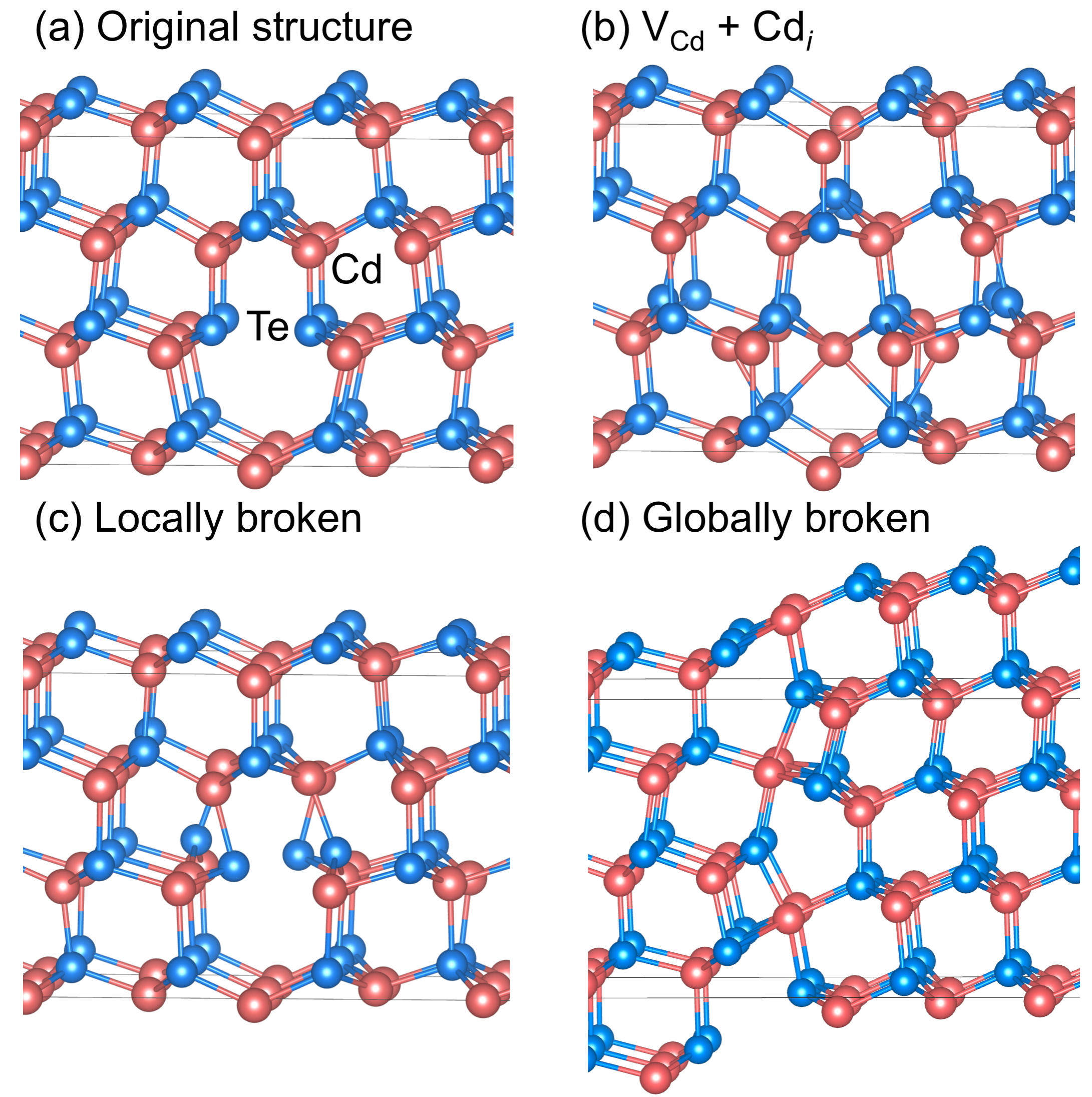}
\caption{\label{fig:2} Atomic structure of some $\Sigma$3 (112) in CdTe, so-called Te core structures. Solid lines represent the cell boundaries.}
\end{figure}

The grain boundary model shown in Figure 2a was initially suggested based on high-resolution TEM images \cite{yan2003structure}.
After a series of studies, it has been suggested that a Cd atom is displaced from the ideal position and form a Frenkel pair (Cd$_i$+V$_{\mathrm{Cd}}$) at the grain boundary for every two unit cells, stabilizing the grain boundary further (Figure 2b) \cite{liu2016self}.
It is also known that the grain boundary can be non-stoichiometric depending on the chemical potentials of the constituents, which can be understood based on the fact that grain boundaries can act as sinks for defects \cite{park2015stability,liu2016self}.
We, however, narrowed our focus on the stoichiometric grain boundaries, and thus the number of atoms is conserved throughout the study unless otherwise mentioned.

While several grain boundary structures were obtained using two levels of computational methods (SIESTA and VASP), we found that the periodicity can be doubled by alternating the dimers of Te atoms, which do not satisfy the Octet-rule (model C), as shown in Figure 2c. 
This structure has almost identical total energy to the originally proposed structure in the PBE calculation, and slightly higher energy in the HSE calculation using the PBE optimized structures (Table 1).
This indicates that thermal vibrations will allow the periodic motions of atoms at the boundary, which can blur the TEM images.
We also successfully obtained a previous model with the Frenkel pairs (model B) \cite{liu2016self}.

The most stable grain boundary model obtained in this study (model D) is shown in Figure 2d, which was obtained by breaking the mirror symmetry globally by rigid body translation.
The atoms were rearranged and one more Cd five-fold coordinated atom and one less three-fold coordinated Te atom are formed in the unit cell as compared to model A (Figure 2a).
This new structure has lower energy of 0.58 eV per dimer than model A.

To examine whether our grain boundary models were thick enough along the boundary normal directions, we added 12 bulk-like layers near the surfaces and compared the total energy of model A and D. The atomic structure of the grain boundaries was not affected by the addition of the layers. We note that we did not obtain the surface energy in this study as we did previously \cite{park2015stability}, therefore the stability of grain boundaries were only investigated by comparing the grain boundary models \cite{park2018quick}.
We obtained the similar relative formation energy of 1.10 eV/nm$^2$ using the PBE functional, indicating that our models are thick enough to obtain the formation energy of grain boundaries.

\begin{figure}
\includegraphics[width=\columnwidth]{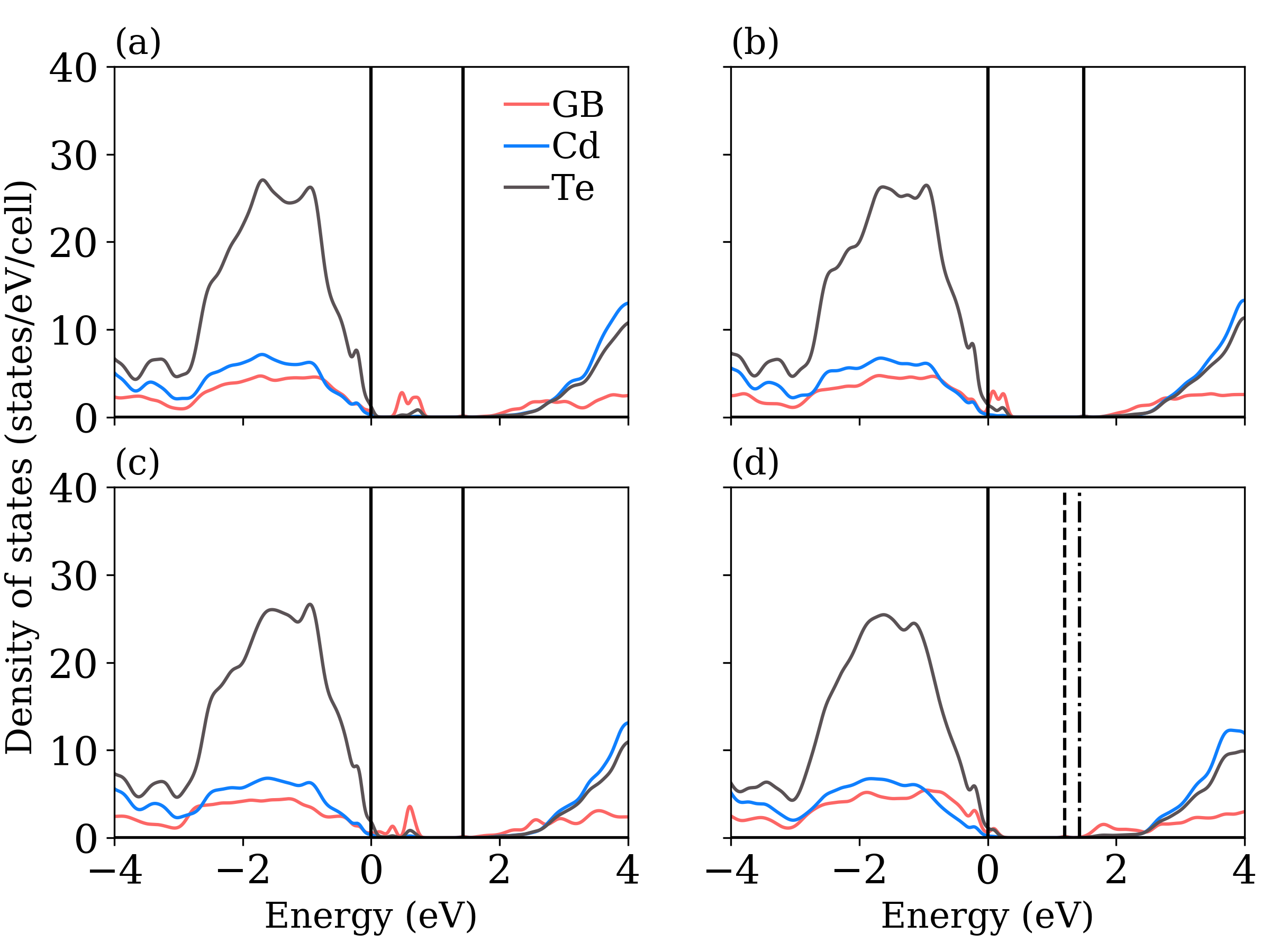}
\caption{\label{fig:3} Electronic structure of the $\Sigma$3 (112) grain boundaries in CdTe calculated by using the HSE06 functional. Each figure shows the projected density of states (PDOS) of the grain boundary model shown in Figure 2. Red, blue, and gray lines represent the PDOS of atoms at the grain boundary, bulk-like Cd, and bulk-like Te atoms, respectively. In each figure, the band edges delocalized in grains are denoted by the solid vertical lines. The energy of the topmost occupied state delocalized in grain(s) is set to 0 eV. In (d), the dashed and dash-dot lines represent the grain boundary states and the estimated conduction band edge position, respectively.}
\end{figure}

The electronic structure of extended defects in compound semiconductors is largely affected by reconstruction and inter-atomic interactions \cite{bennetto1997period,park2015period,park2016polymerization}.
To investigate the effect of the reconstruction on the electronic structure of the grain boundaries, we obtained the electronic density of states (DOS) as summarized in Figure 3.
In the figure, GB stands for the projected density of states (PDOS) of atoms near the boundary. Cd and Te stand for PDOS of remaining Cd or Te atoms in the supercell, respectively.
Consistent with the understanding, our hybrid DFT calculation clearly shows that defect states of grain boundary which were in the middle of the band gap (Figure 3a) became close to the valence band edge by the formation of Frenkel pairs (Figure 3b) and breaking the mirror symmetry (Figure 3d), making the grain boundary less detrimental. 
This result shows that grain boundaries can be self-passivated without impurities and can passivate via mirror symmetry breaking.

\begin{figure}
\includegraphics[width=\columnwidth]{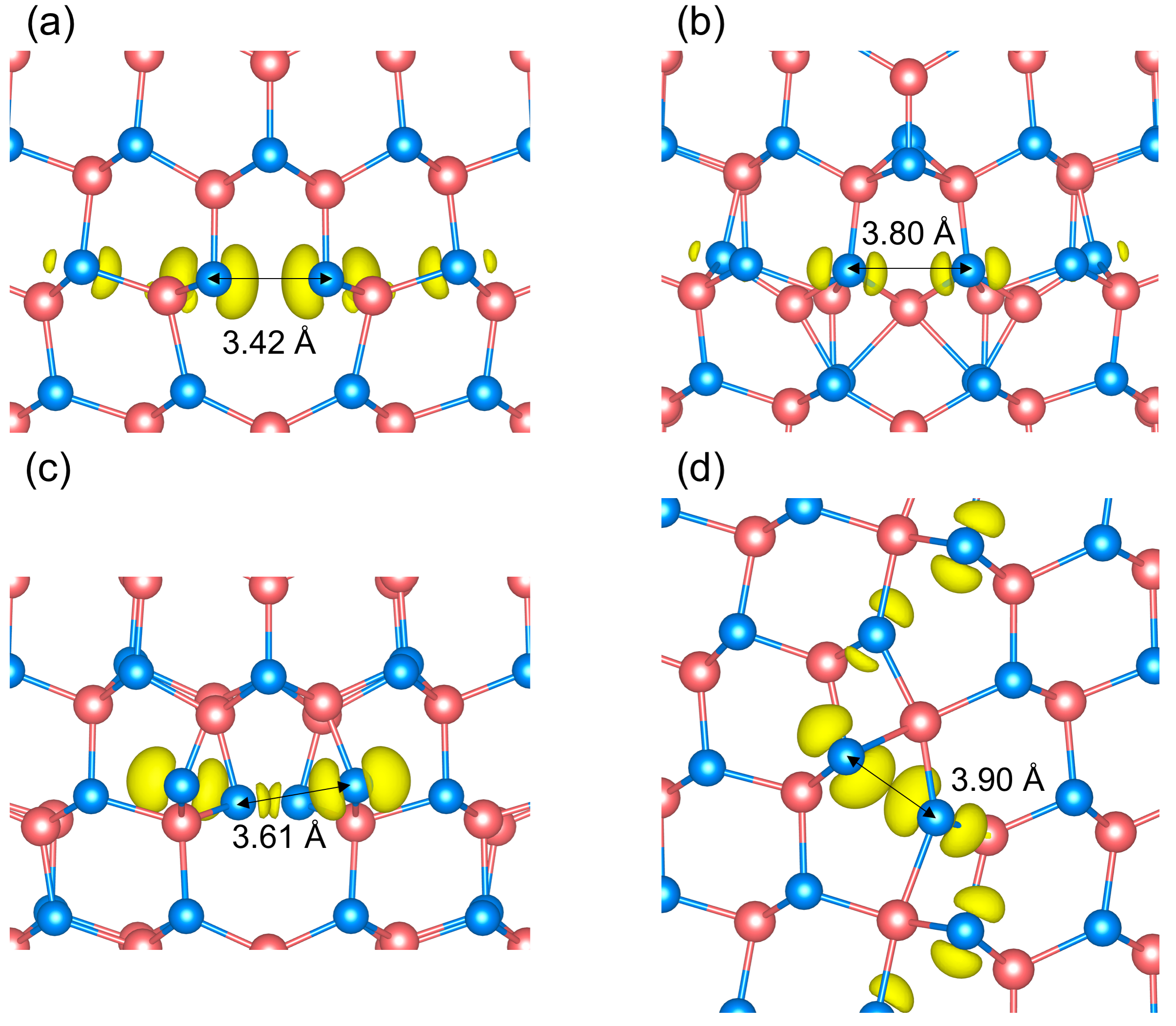}
\caption{\label{fig:4} The occupied gap states composed of Te \textit{p}-orbitals in each grain boundary structure.}
\end{figure}

The less harmful electronic properties of the newly obtained grain boundary originate from the atomic structure of the grain boundary itself.
As shown in Figure 4, the gap states of grain boundaries relatively close to the valence band edge are mostly composed of Te \textit{p}-orbitals.
The Te-Te anti-bonding levels are clearly formed in model A and model C, which have Te dimers with relatively short bond lengths (3.4-3.6 {\AA}).
On the other hand, a Frenkel pair significantly lengthened the distance between the two Te atoms (3.80 {\AA}) in the model B and thus the Te-Te anti-bonding level (Figure 4b) is shifted down in energy.
Our new grain boundary model does not have Te-Te bonds because of the reconstructions and thus it does not have the deep gap states found in model A.
This result is consistent with a previous study showing that the shortened Te-Te bonds generate the anti-bonding above the CBM at $\Sigma$5 grain boundaries  \cite{park2015stability}. 

The study of grain boundaries in CdTe stemmed from the superior performance of solar cells based on polycrystalline CdTe \cite{visoly2004polycrystalline,li2014grain}.
We estimate that the grain boundary model D introduces extended states lower than the CBM of bulk CdTe by about 0.2 eV because of the lengthened Cd-Te bond.
Such extended states of grain boundaries in CdTe might act as channels for electron extraction if charged defects are segregated at the grain boundary and repel hole carriers \cite{li2014grain}.

CdTe is a binary semiconductor without inversion symmetry, and thus we can generate grain boundaries which have Cd dangling bonds by interchanging Cd and Te atoms in the supercell, which is called the Cd core structure (Figure 5a) \cite{yan2003structure}.
This structure was observed less frequently than the Te core structure (Figure 2) in TEM experiments \cite{yan2003structure}. It was later shown that the Te core grain boundary has lower formation energy than the Cd core structure from DFT calculations \cite{park2015stability}.

We also applied the same genetic algorithm method to the Cd core $\Sigma$3 (112) grain boundary and found that the Cd core structure can be further stabilized by breaking the mirror symmetry (Figure 5b and 5c).
When the mirror symmetry is locally broken, every atom is four-fold coordinated and thus the formation energy is lowered by 0.42 eV/nm$^2$ (Figure 5b).
The finding of this structure also demonstrates the difficulties in predicting three-dimensional atomic structure from a two-dimensionally projected image.
When the mirror symmetry is globally broken, five-fold and three-fold coordinated Cd atoms are formed (Figure 5c). This grain boundary model has the lowest formation energy among the Cd-core structures obtained during a series of generations.
The formation energy is further reduced by 0.31 eV/nm$^2$ than the structure with the locally broken mirror symmetry, and thus it is 0.73 eV/nm$^2$ lower than the previously suggested Cd core structure.
Both models have no deep gap states (not shown), indicating that the Cd core grain boundaries are relatively inert.
We also expect that both structures are less stable than model D, based on a previous study reporting the large formation energy difference between the Te core (Figure 2a) and the Cd core structure (Figure 5a) \cite{park2015stability}.

\begin{figure}
\includegraphics[width=\columnwidth]{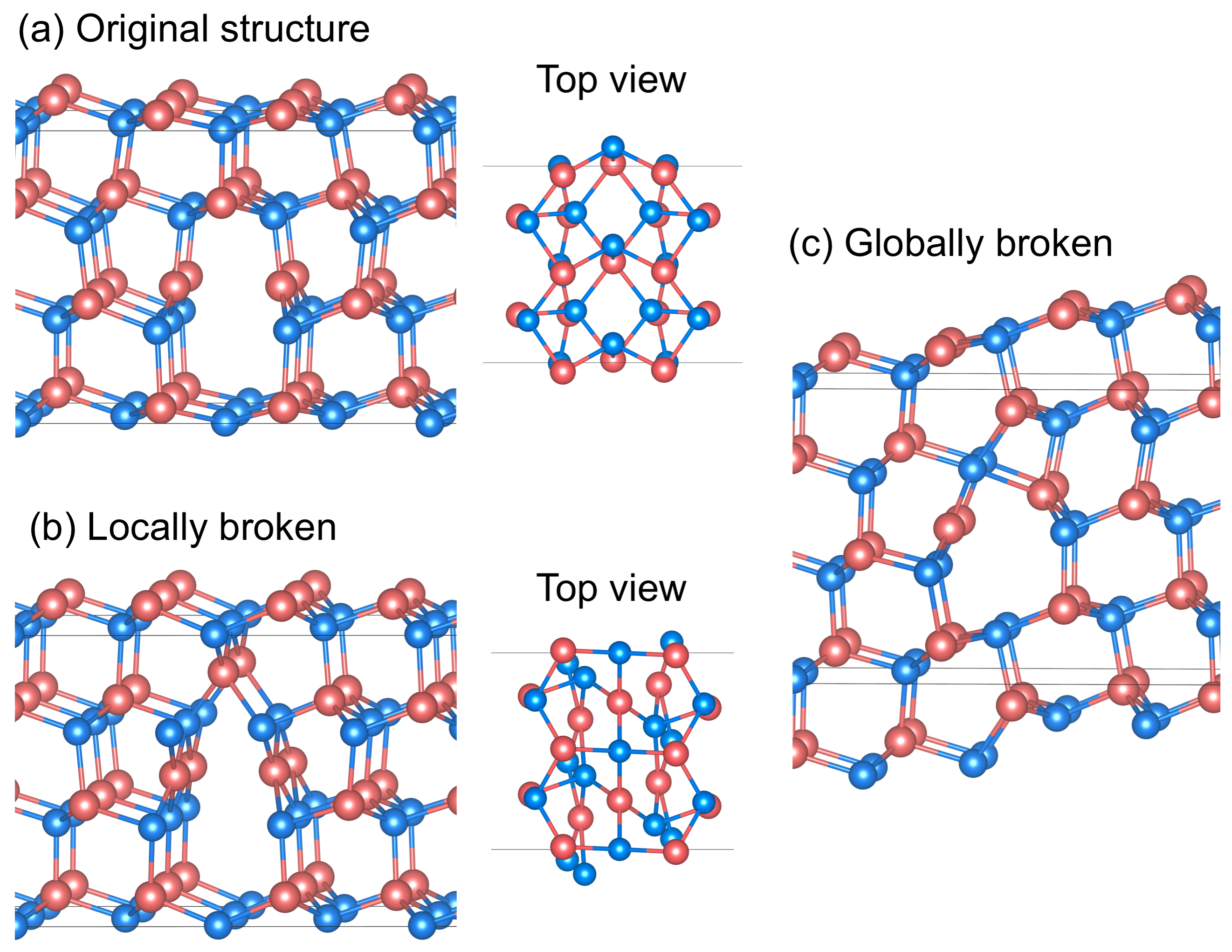}
\caption{\label{fig:5} Atomic structure of some $\Sigma$3 (112) in CdTe by interchanging Cd and Te atoms, so-called Cd core structures. Solid lines represent the cell boundaries.}
\end{figure}

A double positioning twin boundary, which is termed as $\Sigma$3 [112] grain boundary in this study, usually forms together with a lamellar twin boundary \cite{yan2003structure}.
One might expect that the mirror symmetry would not be easily broken globally as the rigid body translation might alter the spacing between layers along the direction normal to the lamellar twin, therefore increasing the strain energy.
We, however, point out that there is experimental evidence of rigid body translation of grain boundaries in silicon \cite{sakaguchi2007atomic,liebscher2018strain} and metals \cite{dehm2002growth,marquis2004finite}.
Even in the previous paper reporting the TEM image of the grain boundary in CdTe, the mirror symmetry seems broken at some layers \cite{yan2003structure}, indicating that the mirror symmetry might be at least partially broken.
A recent classical molecular dynamics simulation also found evidence of small rigid body translation for the Cd core structure \cite{chavez2018molecular}.

Generally speaking, impurities are likely to segregate at grain boundaries, and potentially affect the stable geometry of the grain boundaries.
Cl impurities are particularly important in CdTe solar cells because the solar conversion efficiency is greatly enhanced by the CdCl$_2$ treatment and Cl has been thought to have a significant role on this \cite{li2014grain,yang2016review,park2018point}.
We intentionally introduced a Cl atom in the supercell (Figure 2a) as Cl does segregate at the grain boundaries \cite{li2014grain,yang2016first}.
We attempted to check whether the atomic structure is affected by the incorporated Cl atoms, but we reproduced the stable structure without the Cl atom (Figure 2d). 
This partly justifies the use of the grain boundary atomic structure optimized without impurities for the study of impurities at the grain boundary.

We focused on how to computationally investigate grain boundaries in zinc-blende CdTe, which is probably one of the simplest problems considering the number of chemical elements and symmetry of the host. We, however, believe that our strategy can be easily extended to not only grain boundaries in other polycrystalline materials \cite{yin2014unique} but also other structures likes dislocations \cite{bennetto1997period,park2015period}, interfaces \cite{ono2017theoretical}, and surfaces \cite{merte2017structure,yoo2018selective} by properly restricting the area of interest (e.g. a cylindrical area for dislocations).

Our work presents an efficient method of studying grain boundaries through quantum mechanical simulations. However, our understanding of extended defects in semiconductors is far from comprehensive as compared to that in metals \cite{olmsted2009survey,restrepo2013genetic,yang2018first}. Concepts developed in the investigation of metals should be carefully employed in future studies of grain boundaries in semiconductors. Appropriate methods for dealing with the inherent complexity from multi-components should be considered.

In conclusion, we have proposed a calculation method for finding a stable and/or meta-stable grain boundary atomic structure in materials.
We have found that mirror symmetry of grain boundaries can be locally or entirely broken by periodicity doubling or rigid body translation, respectively.
Grain boundaries in semiconductors like CdTe can be self-passivated by the broken mirror symmetries, which removes the source of the deep levels.
Our new grain boundary model introduces unoccupied extended states below the conduction band of the host, acting as a channel for electron extraction.
We hope that our model may accelerate the study of the atomic structure of these critical features and their impact on device efficiencies in future.

\begin{acknowledgments}
The primary data for this article is available in a repository at \url{https://doi.org/10.5281/zenodo.2369922}.
We thank Dr. Samantha N Hood, Dr. Seung-Sup B. Lee, Prof. Aron Walsh, Prof. Michael Walls, and Dr. Ali Abbas for helpful discussion.
We thanks the Royal Society for a Shooter International Fellowship.
Via our membership of the UK's HPC Materials Chemistry Consortium, which is funded by EPSRC (EP/L000202), this work used the ARCHER UK National Supercomputing Service (http://www.archer.ac.uk). We are grateful to the UK Materials and Molecular Modelling Hub for computational resources, which is partially funded by EPSRC (EP/P020194/1). 
\end{acknowledgments} 

\bibliography{gb.bib}

\end{document}